\documentclass[14pt]{article}

\usepackage[latin1]{inputenc}
\usepackage[T1]{fontenc}
\usepackage[english]{babel}
\usepackage{amsfonts}
\usepackage{amssymb}
\usepackage{amsmath}
\usepackage{amsthm}
\usepackage{graphicx}

\def\be{\begin{equation}}
\def\ee{\end{equation}}
\def\bea{\begin{eqnarray}}
\def\eea{\end{eqnarray}}

\title{Toward a quantitative approach to migrants integration}

\author{Adriano Barra
\footnote{Dipartimento di Fisica, Sapienza Universit\`a di Roma},
Pierluigi Contucci \footnote{Dipartimento di Matematica, Alma
Mater Studiorum Universit\`a di Bologna}}

\begin{document}

\maketitle

\begin{abstract}
Migration phenomena and all the related issues, like integration
of different social groups, are intrinsically complex problems
since they strongly depend on several competitive mechanisms as
economic factors, cultural differences and many others. By
identifying a few essential assumptions, and using the statistical
mechanics of complex systems, we propose a novel quantitative
approach that provides a minimal theory for those phenomena. We
show that the competitive interactions in decision making among a
population of $N$ host citizens and $P$ immigrants, a bi-partite
spin-glass, give rise to a {\it social consciousness} inside the
host community in the sense of the associative memory of neural
networks. The theory leads to a natural quantitative definition of
migrant's ''integration" inside the community.
\newline
From the technical point of view this minimal picture  assumes, as
control parameters, only general notions like strength of the
random interactions, the ratio among the two party sizes and the
cultural influence.
\newline
Few steps forward, toward more refined models, which include some
structure on the random interaction topology (as dilution to avoid
the plain mean field approach) and correlations of experiences
felt among the two parties (biasing the distribution of the
coupling) are discussed at the end, where we show the robustness
of our approach.
\end{abstract}

\section{Introduction}

European Parliament has approved, on November $2008$, the
introduction of a ``blue card'' based on the American green card
counterpart: This issue relies heavily on the introduction of
qualified workers from third countries, putting even more
attention on the subject of ``immigration''.
\newline
Focusing only on EU for simplicity, as USA scenarios are
intrinsically more complex and tangled \cite{USA}, residents of
one member nation of the European Union are allowed to work in
other member nations with little to no restriction on movement.
Due to this policy, traditionally homogenous countries which
usually sent a significant portion of their population overseas,
such as Italy, are seeing an influx of immigrants from several
countries with lower per capita annual earning rates, triggering
nationwide immigration debates \cite{cicero}.
\newline
Barriers to migration come not only in legal form; as a matter of
fact, social opposition to immigration in Europe is visible, the
anti mass immigration perspective is predominantly nationalist and
cultural, rather than economic (at the contrary the latter is
often improved by migration phenomena \cite{card}), conferring to
this issue a social label.
\newline
Again, as a matter of fact, continuous immigration rates persist
and host people is not left unaffected by this: judges with
respect to immigrants are continuously raised as well as debated
inside the community; further, beyond a ''peer-to-peer''
interaction among citizens, media share a certain uniform
influence in orienting popular will on the topic \cite{media}.
\newline
Immigration phenomena, as an aspect of complex social behavior,
surely should match the common properties of complexity
\cite{granovetter}\cite{MPV}, and, as in the whole social science
\cite{white1}\cite{white2}, a lot of attention in  their modelling
has been achieved. Theories have been derived by the so called
``push-pull'' models \cite{euro}, or based on dynamical systems
\cite{buckley}\cite{lee} (i.e. partial differential equations
usually involving time), but never the scenario has been plugged
into a statistical mechanics framework. In this work we want to
attempt a first minimal step toward a quantitative approach in
this direction.
\newline
Disordered statistical mechanics, even though developed in the
framework of theoretical physics \cite{barabasi}, deals
(prevalently) with the equilibrium properties of collections of
agents which interact among them via competitive exchanges
\cite{MPV}: it reveled a surprisingly high capability to manage
field of research far form physics as neurobiology \cite{amit},
immunology \cite{parisi}, economics \cite{bouchaud} and, recently,
quantitative sociology
\cite{brock}\cite{contucci1}\cite{contucci2}\cite{marsili}.
\newline
The main idea is that when the amount of agents is sufficiently large,
disordered statistical mechanics comes in help as a stochastic
optimization procedure which  aims to predict the global behavior
of complex systems by considering, in a probabilistic
framework, the microscopic deterministic dynamics of all the
constituents of the system itself.
\newline
Its methodical application for building a ''quantitative
sociology'' has been strongly advocated by social scientists with
a background in microeconomics \cite{durlauf}.
\newline
As a first attempt we propose a simple
model for analyzing equilibrium decision making \cite{macfadden}
when a population of individuals (i.e. a country) experiences
interactions with incoming people afferent by several other places
all over the world.
\newline
As customary within theoretical physics methodologies, the model
has a certain number of ''assumptions'' that the reader may find
''too strong to describe reality''. This is a key feature of this
 approach: dealing with all the details of the phenomenon is surely
appealing but more than prohibitive for the complexity involved.
Moreover there is the danger of making a {\em snapshot} of the
reality rather than a model with some predictive and descriptive
capabilities. On the contrary, if via a minimal model some
features can still be retained, this may offer a suggestion for
understanding the main degrees of freedom to take into account for
future speculations.
\newline
In the last section we will show that even (partially)
relaxing our simplifying assumptions (that we are going to emphasize here
after), the features of our model are retained.
\newline
To simplify the mathematical treatment at this first stage, what
we postulate is:
\begin{itemize}

\item When making the host and the migrant communities
interact, we neglect the underlying social network (the so called
``small world'' \cite{smallworld}) such that each of the $N$
citizens, sooner or later, meet each of the $P$ migrants (mean
field approximation).

\item Noticing that the amount of the host agents is huge, many
of them will share a negative experience with the migrants while
many others will share a positive one; these ``exchange
interactions'' $\xi$ are thought of to be randomly chosen with a
centered symmetric probability, i.e. $\mathbb{P}(\xi = +1) =
\mathbb{P}(\xi = - 1) = 1/2$.

\item Every dependence of the host opinions on the rate of
migration is also refused (leaving this task for future
improvements of the model) and we consider only the amount of
foreigners with respect the host bulk.

\item Cultural background (including media influence on the host community)
that we encode via an external field $h$ is taken into account to
influence uniformly each citizen; models with random or
time-dependent fields are again left for future improvements.

\end{itemize}

While, at least the limits of $P/N \to 0$ (no immigrants scenario)
and $P/N \to \infty$ (only immigrants scenario) are clear and
poorly interesting as they correspond to absence of social
mixtures, in between those two limits complex behaviors may occur.
\newline
We obtain two (intertwined) main results:
\begin{itemize}

\item First we prove
 that, when considering our model for interactions among these two parties, this is
equivalent (shares the same equilibrium features) to considering a
particular Hopfield model of associative neural network
\cite{barraguerra} inside the host community: In ``social terms'',
when looking at the host country, the population spontaneously
develops a ``consciousness'' of the interactions among the
immigrants. Experiences of the migrants are stored  into a
collective ``social memory''.

\item Interestingly, this happens if the ratio among the amounts of
immigrants over the hosts is smaller than a few per cent
(depending on the particular choice of the control parameters).
\newline
If the ratio becomes higher (toward an high multi-ethnic picture)
this property falls down (blackout scenario \cite{amit}) and there
is no memory of the past interactions: the system ages as a
canonical spin-glass \cite{MPV}, defining a transition toward an
integrated state of the immigrants by the host community.
\end{itemize}
We stress that our model does not predict any transitions from an
hostile attitude toward a benevolent one; instead  there is a
phase in which the host collective memory is able to formulate
opinions on the migrants (these can be both positive or negative
or even fluctuating) and there is a phase in which this memory
does not work any longer (the immigrants become unaffected by the
judge). In the latter, they lose their ''immigrant'' label and
indifference with respect to a judge in this sense is achieved,
which we define as {\em social integration} or better {\em
normalization} in the sense that the foreigners presence has
become normal and is no longer noticed.

\section{Social memory of immigrant's interactions}

Let us concretely consider a first population of $N$ citizens $i
\in (1,...,N)$; each citizen has an opinions $\sigma_i \in \pm 1$,
where $+1$ represents a positive  attitude with respect to the
migrants, viceversa for $-1$. As this is supposed to be the host
population ensemble, no null judges are allowed for the citizens
($\sigma \neq 0$): each agent is suppose to share a cultural
relation with the others and must take a net position with respect
to the phenomenon (the diluted case allowing even $\sigma=0$ will
be briefly discussed at the end).
\newline
The second population is made by $P$ real agents $\tau_{\mu} \in
\mathbb{R}, \mu = 1,...,p$. This community is composed by all the
incoming immigrants and, to emphasize this lacking of cultural
aggregation, its a-priori probability distribution $\mathbb{P}$ is
chosen as a standard centered Gaussian, i.e.
\begin{equation}
\mathbb{P}(\sigma_i) = \frac12 \large( \delta(\sigma_i - 1) +
\delta(\sigma_i + 1) \large), \ \ \mathbb{P}(\tau_{\mu}) =
\frac{1}{\sqrt{2 \pi}}\exp\big(-\tau^2_{\mu}/2\big).
\end{equation}
Both the communities need to be very large, as, dealing with phase
transitions \cite{ellis}, we will be interested in making both $N$
and $P$ diverge, such that their ratio is held finite and acts as
a control parameter of the theory, i.e. $0 < P/N=\alpha \ll 1$
(for example in Italy the amount of citizens is $\sim 60 \cdot
10^6$ and the amount of immigrants $\sim 30 \cdot 10^5$, both are
''large numbers'' and their ratio defines $\alpha \sim 5 \cdot
10^{-2}$ ).
\newline
Another control parameter is $\beta$ which tunes the strength of
the interactions among the two parties (it can also be thought of
as the inverse of a noise inside the interaction network
\cite{barraguerra}, reflecting the lacking of a full deterministic
behavior as depicted by our assumption).
\newline
Now we encode into a cost function (an Hamiltonian
$H_{N,P}(\sigma,\tau;\xi,h)$, where $\xi$ is quenched and $h$
externally tunable) these interactions in a mean field way, via
both positive and negative couplings $\xi_{i}^{\mu}$ such that
\be\label{latre} H(\sigma,\tau;\xi,h) =
-\frac{1}{\alpha^{1/4}\sqrt{N}}\sum_{i,\mu}^{N,P}\sigma_i
\xi^i_{\mu} \tau_{\mu} - h \sum_i^N \sigma_i, \ee with
$\mathbb{P}(\xi_i^{\mu}=1)=\mathbb{P}(\xi_i^{\mu}=-1)=1/2$,
($\mathbb{P}$ being the $\xi$ probability distribution).
\newline
The meaning of the cost function (\ref{latre}) is clear: each
citizen $i^{th}$ meets each migrant $\mu^{th}$, further,
experiences a {\em pressure} by the media and his historical
background, which influence uniformly the host opinions via the
external field $h$.
\newline
The success of the meeting is suggested (but, due to the noise
$\beta$, not strictly imposed) by the value of the $\xi_i^{\mu}$:
if $\xi_i^{\mu}>0$ both the $i^{th}$ citizen and the $\mu^{th}$
migrant will tend to have the same reciprocal judgement (i.e.
$\sigma_i > 0, \ \tau_{\mu} > 0$ or $\sigma_i <0, \ \tau_{\mu} <
0$): It can be positive (for example a low cost employee offer
which may satisfy both of them) or negative (for example a robbery
which may give rise to reciprocal hostility). $\xi_i^{\mu}<0$ is
also clear: it reflects cases where only one of the two meeting
agents is satisfied by the interaction while the other do not.
\newline
Overall the set of obtainable scenarios for interactions is
complete.
\newline
We stress that the single event is not of fundamental importance:
we are interested in understanding how the country globally
responds to the phenomenon. Further, for a particularly bad
situation, given the huge number of agents, there will be probably
a particularly good one and even more important a judgement in a
single citizen will probably change over the time, but as one may
change, even the others, leaving the averaged opinion unaffected
by these fluctuations.
\newline
The global country judgement on the migrants is defined obviously
as $m=N^{-1}\sum_i^N \sigma_i \in \{-1,+1\}$ such that if $m=-1$ a
complete antagonist attitude is shown, (and viceversa for $m=+1$),
while $m=0$ stands for indifference.
\newline
The square root in the normalization of the Hamiltonian (instead
of a power $1$), reflects the several cancelations which happens
on the volume of the agents due to the conflictingly interactions
\cite{barra0}.
\newline
Once introduced the partition function $Z(\alpha,\beta;\xi,h)$,
defined as \be Z(\alpha,\beta;\xi,h)=
\sum_{\{\sigma\}}\int_{-\infty}^{+\infty} \prod_{\mu}^P d\tau_\mu
e^{-\tau_{\mu}^2/2}\exp\Big(-\beta H(\sigma,\tau;\xi,h)\Big),\ee
and using $\langle \rangle_{\xi}$ as the average over the quenched
$\xi$ distribution, the equilibrium properties of this system are
found by studying the related free energy $f(\alpha,\beta,h)=
(-1/\sqrt{\alpha}N)\langle \ln Z
(\alpha,\beta,\xi,h)\rangle_{\xi}$.
\newline
In a nutshell, at each given triplet $(\alpha,\beta,h)$, the
minima of this free energy display equilibrium among cost function
minimization and severe entropy related constraints, as, due to
the dichotomy of the interactions, the system behaves as a
bipartite spin-glass \cite{barraguerra}.
\newline
By applying the Hubbard-Stratonovich lemma \cite{ellis} on the
Gaussian fields $\tau$, we map the partition function
$Z(\alpha,\beta;\xi,h)$ to \be Z(\alpha,\beta;\xi,h)=
\sum_{\{\sigma \}} e^{- h \sum_i^N \sigma_i}
e^{-\frac{\beta^2}{\sqrt{\alpha}N}\sum_{i,j}^N(\sum_{\mu}^p
\xi_i^{\mu}\xi_j^{\mu})\sigma_i \sigma_j}. \ee A new cost function
 $\tilde{H}(\sigma; \xi,h)$, in terms of only agents $\sigma$
 interacting each another, is obtained as \be\label{hop} \tilde{H}(\sigma;\xi,h) =
\frac{-1}{\sqrt{\alpha}N}\sum_{i,j}^{N,N}(\sum_{\mu}^P
\xi_i^{\mu}\xi_j^{\mu})\sigma_i \sigma_j + h\sum_i^N \sigma_i. \ee
This second Hamiltonian (an Hopfield model
\cite{AGS2}\cite{hopfield}) is {\em hidden} into the first, it
shares with it the same equilibrium behavior (i.e. the same value
of $m$)
 and accounts for {\em social consciousness}.
\newline
What it encodes is simple: if both $i^{th}$ and $j^{th}$ citizens
had a good (or bad) interaction with respect to the $\mu^{th}$
immigrant
 ($\xi_i^{\mu}=\xi_j^{\mu}$), then, when they meet (i.e. through
the coupling $\sigma_i\sigma_j$), $\xi_i^{\mu}\xi_j^{\mu}$ is
positive and they reinforce their opinion on the $\mu^{th}$
immigrant.  At the contrary, if $i^{th}$ has a good interaction
with the migrant and $j^{th}$ a bad one (or viceversa)
$(\xi_i^{\mu} \neq \xi_j^{\mu})$, the exchange among the two
citizens weakens their viewpoints.
\newline
To have the global behavior one must then sum over all the citizen
couples and all the migrants.
\newline
We note that the interaction strength among the same party (i.e.
citizen with citizen) is quadratically stronger than with the
immigrants (host community trusts much more on the judgement of
its elements).
\newline
So we see that the  model we introduced behaves, in the space of
the only citizens, as an associative memory \cite{amit} with
Hebbian synapses i.e. $J_{ij} = N^{-1}\sum_{\mu}^p
\xi_i^{\mu}\xi_{j}^{\mu}$ \cite{hebb}\cite{hopfield}.
\newline
The mapping is in fact robust: The $N$ dichotomic host judgements
$\sigma_i$ of the social network reflect the $N$ dichotomic
(quiescent or firing) neurons of the neural network.
\newline
Further, the dichotomic felt experiences $\xi_i^{\mu}$ in social
network map the dichotomic ''learned'' experiences  in neural
network,  and the media influence over the host community $h$
mirrors the threshold for firing \cite{amit}; even the power one
into the normalization of the two body interaction in eq.
(\ref{hop}) turns out to be the proper one of neural networks
\cite{barraguerra}.

\section{Migrant's integration inside the host community}

This bridge is extremely interesting as it naturally pioneers a
quantitative definition of {\em social integration} for the
migrants. In fact a long and deep debate among scientists involved
in social research is  still going on the choices of the
quantifiers for this phenomenon \cite{pipino}, ranging from
averaged hire equivalence among citizens and migrants to the
percentage of shared marriages with respect to inner community
ones, as well as several others possible indicators \cite{portes}.
\newline
In our framework, the interactions among the two parties are
stored as memories into the dialogues of the host population such
that the network's free energy has $P$ minima, each of which
corresponding to a given $\xi$ (the gauge symmetry which would
suggest $2P$ minima is broken by the field $h$).
\newline
Let us now focus on the memory capacities of the model: Starting
from a generic point of the $2^N$ possible states of the system
made by the citizens, and propagating through a meaningful
Markovian dynamics (i.e. a Glauber prescription \cite{amit} as the
cost function (\ref{hop}) obeys detailed balance \cite{ellis}),
the free energy collapses into the minimum of the closer basin of
attraction, which are the $P$ learned patterns or their linear
combinations (spurious states).
\newline
A convenient order parameter set to check the ''retrieval'' of the
network is the $P$-vector Mattis magnetization $m^{\mu} =
N^{-1}\sum_i^N \sigma_i \xi_i^{\mu}$, $\mu \in (1,...,P)$
\cite{amit}, such that, if the control parameters range in the
associative memory phase, there will be at least a $\mu$ for which
$m_{\mu}\to1$ after enough time.
\newline
Of course increasing $P$ over a threshold, the Hebb matrix will
approach a Gaussian for the Central Limit Theorem (blackout
scenario): when this happens the neural networks turns into a
spin-glass (a Sherrington-Kirkpatrick model
\cite{barra1}\cite{guerra}), the amount of minima becomes
exponentially proportional to the bulk \cite{MPV}  (and not
linearly, i.e. $P \sim \alpha N$), the basins of attraction become
all condensed one into another and retrieval is no longer
possible.
\newline
So, if elapsing time, the amount of migrants increases (as well as
 the number of host experiences) up to this threshold $\alpha_c
(\beta,h)$, that defines the blackout: the transition to a glassy
landscape makes retrieval no longer achievable and, in the host
social network, it is not possible to formulate a global statement
of the phenomenon, i.e. $m=0, m_{\mu}=0 \forall \mu \in
(1,...,P)$: Consequently the immigrants are integrated as they can
no longer be thought of as ''immigrants''.
\newline
To continue with quantitative results, let us discuss the network
still in the zero noise regime ($\beta \to \infty$) and with an
infinitesimal field $h$ that assures only gauge breaking, while
for general scenario we remind again to textbooks specialized in
neural networks \cite{amit}\cite{peter}.
\begin{itemize}
\item When $\alpha<5\%$ there is coexistence of retrieval and
spin-glass phases, with retrieval minima lower with respect to the
spin-glass ones. $m_{\mu} \neq 0$ for sparse values of $\mu$ and
spurious states play a considerable role (see the next section).

\item When $5\%<\alpha<14\%$ there is coexistence of the retrieval
and spin-glass phases, with spin-glass minima lower with respect
to the retrieval ones. No spurious states exist and only one
pattern each time can be minded.

\item When $\alpha>14\%$ only the spin-glass phase survives. There
is no longer any retrieval and the {\em social consciousness} on
the phenomenon disappeared.
\end{itemize}

At the end we want to report some extensions relaxing some of the
hypotheses we followed. It is well known in social theory that the
mean field approach we used is too rude. In fact, from the early
investigations of Stanley Milgram \cite{milgram} up to the recent
formulations by Watts and Strogatz \cite{smallworld}\cite{SW1}, it
is known that a proper underlying topology for the interactions is
the {\em small world}. It has been shown by Coolen and coworkers
that neural networks on these graphs are very close to the
standard one by Hopfield \cite{ton1}\cite{ton2}.
\newline
Furthermore the fully uncorrelated experiences maybe a too rude
simplification as well. It is in fact very natural to assume that
correlations, at least due to similarity among migrants with
several common features (i.e. religion, political views,
lifestyle), do exist. To take into account these features we
should introduce a probability distribution for the experiences as
$$
P(\xi)=\frac12 (1+a) \delta(\xi+1)+ \frac12 (1-a) \delta(\xi-1),
$$
which naturally enlarges the previous scheme (that is recovered
here for $a=0$) as $\langle \xi_i^{\mu} \xi_i^{\nu}\rangle =
\delta_{\mu \nu} + a^2(1-\delta_{\mu \nu})$. In \cite{AGS2}, Amit
and coworkers developed a correlated memory neural network in
which they showed that upon rescaling the Hebbian synaptic matrix
as
$$
J_{ij} = \sum_{\mu}^{p}(\xi_i^{\mu}-a)(\xi_j^{\mu}-a)
$$
again the behavior of the network is largely unaffected by the
modification, the whole conferring a certain degree of robustness
to our minima model.

\section*{Acknowledgments} The authors are grateful to Marzio
Barbagli for useful stimulating discussions. AB is indebted with
Francesco Guerra, with which he started a methodical analysis both
of neural networks and spin glasses. AB research is supported by
the Smart-Life Project which is acknowledged as well as a grant by
GNFN for mobility. PC acknowledges partial financial support from
Strategic Research Grant of University of Bologna.

\end{document}